\documentclass[USenglish,oneside,twocolumn]{article}

\usepackage[utf8]{inputenc}
\usepackage[big]{dgruyter_NEW}
\usepackage{array}
\usepackage{dingbat}

\begin{document}

  \author*[1]{Parisa Kaghazgaran}

  \author[2]{Hassan Takabi}

  \affil[1]{Texas A\& M University, kaghazgaran@tamu.edu}

  \affil[2]{University of North Texas, Hassan.Takabi@unt.edu }

  \title{\huge Privacy-preserving Edit Distance on Genomic Data}

  \runningtitle{ESCOT}


  \begin{abstract}
{Suppose Alice holds a DNA sequence and Bob owns a database of DNA sequences. They want to determine whether there is a match for the Alice's input in the Bob's database for any purpose such as diagnosis of Alice's disease. However, Alice does not want to reveal her DNA pattern to Bob, since it would enable him to learn private information about her. For the similar reasons, Bob does not want to reveal any information about his database to Alice. This problem has attracted attention from bioinformatics community in order to protect privacy of users and several solutions have been proposed. Efficiency is always a bottleneck in cryptography domain. In this paper, we propose \textbf{\textit{ESCOT}} protocol to address privacy preserving Edit distance using Oblivious Transfer (OT) for the first time. We evaluate our approach on a genome dataset over both LAN and WAN network. Experimental results confirm feasibility of our approach in real-world scenarios.}
\end{abstract}



\maketitle

\section{Introduction}
Edit distance algorithm approximates how similar two DNA sequences are. This similarity then can be used in finding similar cancer patients across organizations which helps in deciding the appropriate treatment. Due to privacy constraints, organizations are not willing to reveal their clients' private information. 

\textbf{Problem Definition:} In this paper, we aim to propose an efficient privacy preserving protocol to find the most similar patients in a database on a panel of genes measured by the Edit distance between a query sequence and sequences in the database. Formally, the server and client want to privately compute the distance between one versus $N$ sequences. The client's input represents a sequence $X$ while the server holds $N$ sequences {$Y_1,..., Y_N$}. Then, they jointly calculate the Edit distance between $X$ and each $Y_i$. At the end of the protocol, $Y_i$s with less than $k$ distance to $X$ will be returned to the client.

This problem is a type of secure two-party computation in which two parties jointly compute a function on their private inputs without disclosing their data to each other except the final output. Recently, secure computation has attracted attention in computational biology and bioinformatics to preserve privacy of biological data  \cite{cetin2017private, kim2017secure, barni2010privacy, blanton2011secure, evans2011efficient, chun2014outsourceable}. Early approaches were based on pure (additive) Homomorphic Encryption (HE) e.g., \cite{erkin2009privacy}. Later work showed that protocols using generic secure computation techniques such as Yao's garbled circuits and GMW circuits outperform HE. These protocols are based on either a combination of HE and circuit-based approaches \cite{barni2010privacy, blanton2011secure, evans2011efficient} or pure circuit-based techniques \cite{bringer2012faster, huang2011faster, luo2012efficient}.

On the other hand, recent optimization of Oblivious Transfer (OT) -- known as OT extension--  is recognised as the most efficient technology in \textbf{two-party} settings \cite{asharov2013more, kolesnikov2013improved, demmler2015aby}. For example, Demmler et al. showed that an OT-based solution for privacy-preserving identification outperforms HE and circuit-based techniques \cite{demmler2015aby}. To the best of our knowledge, we propose a privacy preserving Edit distance using OT extension for the first time. 

However, communication is a bottleneck in the existing OT based protocols. Schneider showed that an increase in the bit-length of the transferred data and/ or an increase in the number of the required OT lead to a significant increase in communication bandwidth \cite{BAYSlides}. Kolensnikov et al. improved OT extension for transferring short messages such as binary messages \cite{kolesnikov2013improved} to address communication overhead.

As comparison of two values is the main building block for Edit distance algorithm, we reduce the problem of secure Edit distance into secure comparison and propose an \underline{E}fficient \underline{S}ecure \underline{C}omparison protocol based on \underline{O}blivious \underline{T}ransfer (\textbf{\textit{ESCOT}}) with binary representation of the data transferring among computing parties. We also provide the security and accuracy analysis of \textbf{\textit{ESCOT}}. We implemented \textbf{\textit{ESCOT}} algorithm in Java and the source code will be available  to reproduce the results by the time the paper is published. 

\section{Background}
In this Section, we provide some background information about Edit distance algorithm and Oblivious Transfer (OT).
\subsection{Edit Distance}
\textbf{Definition:} Edit distance $d(X, Y)$ between two strings or sequence of characters $X, Y$ is the minimum number of insertions, deletions and substitutions required to transform $X$ into $Y$. 

In classic distance measures like Euclidean distance, the sequences are required to have equal length. Therefore, for measuring the similarity between DNA sequences with different lengths, more complicated distance measures like Edit distance are needed. 

In our scenario, the client owns $m$-length sequence of characters $X$ and the server has $m'$-length sequence of characters $Y$ where $m \neq m'$ and the characters belong to a finite alphabet set. $X$ can be transformed into $Y$ by applying Edit distance algorithm. The Edit distance is the minimum aggregate cost necessary to perform this transformation. 

The basic Edit distance algorithm is Wanger-Fisher \cite{wagner1975complexity} which works in a brute-force manner to compare two sequences and its complexity is $O(m\times m')$ or $O(m^2)$. This quadratic complexity is inefficient in the cryptography domain. Therefore, we need to seek for more sophisticated algorithms with lower complexity that help privacy preserving genome analysis.

Ukkonen's algorithm \cite{ukkonen1985algorithms} improves  Wagner-Fisher algorithm by limiting the number of operations, provided that the Edit distance is less than a given threshold $k$. It runs in $O(m \times k)$ time which improves the complexity significantly when the sequences are lengthy. 

\subsection{Oblivious Transfer (OT)}\label{OT}
In Oblivious Transfer (OT), two parties known as sender and receiver participate in protocol. In 1-out-of-2 OT, sender has two private messages ($x_0$, $x_1$) and receiver has a selection bit   $r \in \{0, 1\}$. At the end of the protocol, the receiver only learns $x_r$ and learns no information about $x_{1-r}$ and the sender learns nothing about $r$.  In its generalized form i.e., 1-out-of-n OT, the sender has $n$ messages $\{x_0,...,x_{n-1}\}$ while the receiver has a selection value $r\in\{0,...,n-1\}$ to obtain $x_r$.

Preliminary OT-based protocols consist of expensive public-key operations while recent improvements of OT, called OT-extension \cite{asharov2013more, kolesnikov2013improved}, allow  executing many OTs using only symmetric operations with a constant and small number ($\kappa$) of public-key operations as base-OTs.

To perform base OTs, we can use either homomorphic encryption or Deffie-Hellman key exchange protocol \cite{chou2015simplest}. In our implementation, we use the latter one as follows: 

The sender selects a random number $a \in Z_p$ and sends $A=g^a$ to the receiver ($g$ is the group generator). The receiver picks a random number $b \in Z_p$ and calculates $B=g^b$ (if $r=0$) or $B=Ag^b$ (if $r=1$). He then sends $B$ to the sender. The sender calculates $k_0=H(B^a)$ and $k_1= H((B/A)^a)$ such that $k_0$ and $k_1$ act as the secret keys in a symmetric encryption $E$. Then, he encrypts its messages, $e_0=E_{k_0}(x_0)$ , $e_1=E_{k_1}(x_1)$, and sends the ciphertexts to the receiver. Then, the receiver calculates $k_R= H(A^b)$ and decrypts the desired message by its key as $x_r=D_{k_R}(e_r)$. $H$ stands for a secure hash function.
  
The security of base OT protocols directly depends on the security of the underlying homomorphic encryption or Deffie-Hellman protocols. In particular, Deffie-Hellman  key exchange protocol is secure due to the hardness of computing discrete logarithms. Now, the results obtained from the execution of the base-OT are used to perform many OTs efficiently using lightweight symmetric operations.

The OT extension protocol proposed in \cite{kolesnikov2013improved} is the recent optimization of OT extensions, specially for short messages, that supports 1-out-of-n ($n > 2$) OT in addition to 1-out-of-2 OT. We adopt this algorithm in our secure comparison protocol. 
 \\
 \\
\section{The Proposed Approach}
In this section, we first propose our secure comparison protocol. Then, introduce our algorithm for privacy preserving Edit distance. 

\subsection{ESCOT Protocol}
\begin{figure}
  \label{algED}
  \centering
    \includegraphics[width=0.5\textwidth]{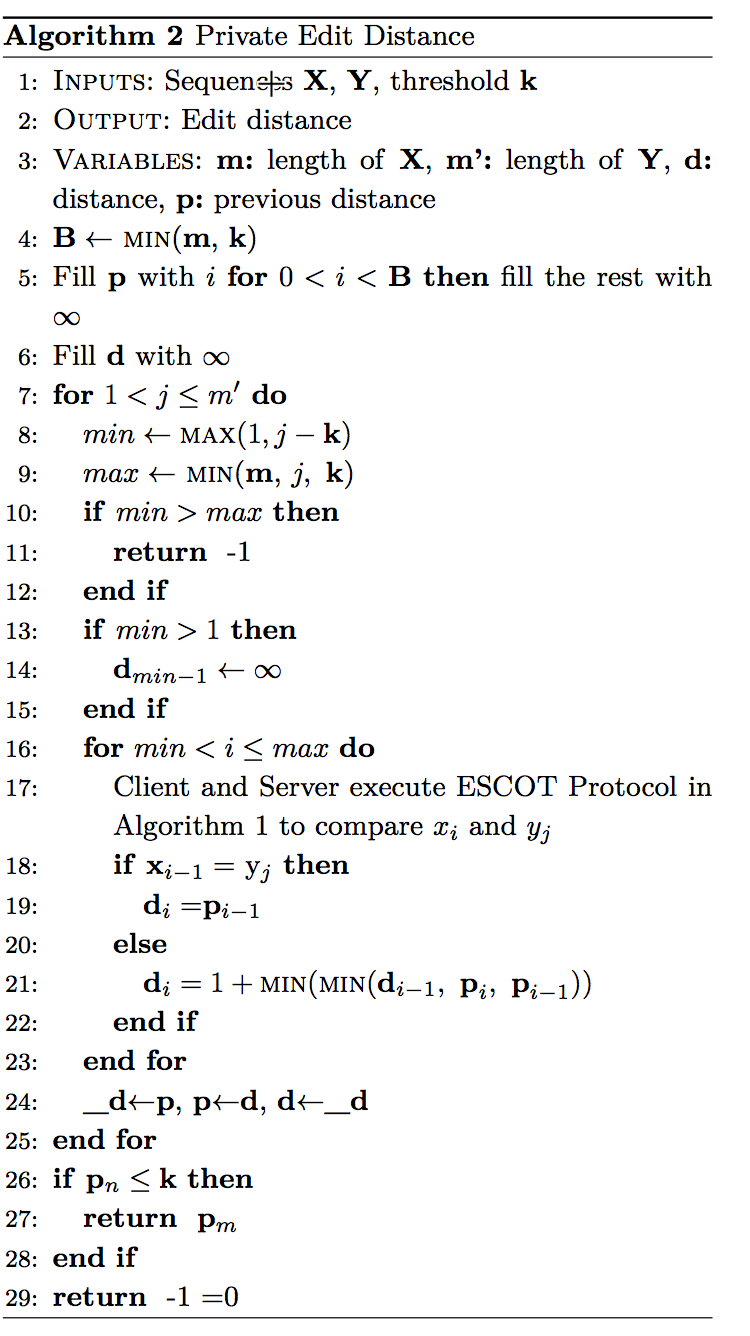}
\end{figure}

Unlike classic distance measures like Euclidean distance which are composed of four basic mathematical operations ($+, -, \times, \div$), Edit distance is based on boolean comparison. It checks whether two specific characters from two separate sequences are equal (Algorithm \ref{algED}, Line 18). 

We propose a novel protocol for secure comparison based on OT called \textbf{\textit{ESCOT}}. For example, $X=\{x_0,...,x_{m-1}\}$ and $Y=\{y_0,..., y_{m'-1}\}$ are respectively the client and server's sequences in an alphabet set of size
$n$. If we encode the characters as numbers, the code value vary from $0$ to $n-1$. The goal of secure comparison is to check whether $x_i$ and $y_j$ are equal where $0\le x_i, y_j \le n-1$.

In \textbf{\textit{ESCOT}} protocol, client acts as the receiver and server acts as the sender in OT.  Sender generates $n$ number of OT messages for each character $y_j$ in its sequence as follows:

\begin{equation}
 M_{k (0\le k \le n-1)}= \left\{\begin{array}{lll}
            1 & if & (k-y_j)\  mod\  n=0\\
            0 & else \\
            \end{array}\right.
\end{equation}

On the other side, receiver puts the value of $x_i$ as its selection bit. Since the number of OT messages is $n$, execution of 1-out-of-n OT is required. The logic of this protocol is that if $x_i$ and $y_j$ are the same, then $1$ will be transferred; otherwise, $0$ will be transferred. Since the length of OT messages is one bit, execution of the 1-out-of-n protocol proposed in \cite{kolesnikov2013improved} is highly fit for our problem as it is the most efficient protocol known today for short-length messages. \textbf{\textit{ESCOT}} protocol to compare $x_i$ and $y_j$ is described in Algorithm 1.

\begin{figure}
  \label{ESCOT-alg}
  \centering
    \includegraphics[width=0.5\textwidth]{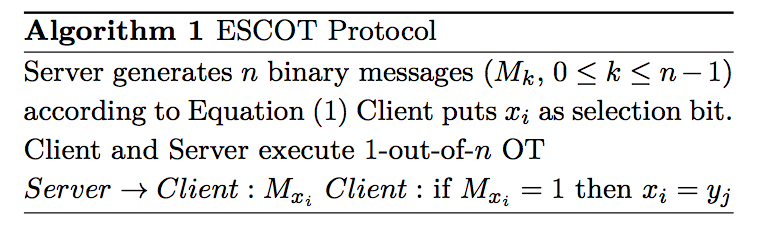}
\end{figure}

\textbf{Correctness Analysis:} The message corresponding to the selection bit is transferred ($M_{x_i}$). Intuitively, if $x_i=y_j$ then the condition $(k-y_j)$ $mod$ $N=0$ is satisfied and the message is 1. If $x_i<>y_j$ then the transferred message is 0.

\textbf{Security Analysis}: The security of \textbf{\textit{ ESCOT}} depends on security of the underlying OT protocol. The receiver only receives the message corresponding to its selection bit and gain no information about the other messages. The sender will not learn any thing about the selection bit. In addition, in the Edit distance algorithm we consider in this paper, only sequences with enough similarity (based on the threshold $k$) are processed to the end and if the Edit distance exceeds the threshold then the execution will stop. This way, we can minimize the information leakage.

\textbf{Communication Analysis}: For each comparison, the communication bandwidth takes $\kappa+n$ bits where $\kappa$ is security parameter. Edit distance algorithm requires $k \times m$ number of comparisons so the consumption of the bandwidth is $(k \times m)\times (\kappa + n)$ bits in our algorithm.

\subsection{Private Edit Distance based on ESCOT:}

In Algorithm 2, we combine Ukkonen's Edit distance algorithm \cite{ukkonen1985algorithms} with our \textbf{\textit{ESCOT}} protocol to address privacy preserving Edit distance.

\section{Experiments}\label{ED protocol}
To calculate Edit distance between two sequences, \textbf{\textit{ESCOT}} protocol executes $k \times m$ times, where $k$ is the distance threshold and $m$ is the maximum length of the input sequences.

\textbf{Threat Model:}
Our threat model is semi-honest that means both sender and receiver follow the protocol specification accurately and do not try to change their messages in order to obtain private information from other party. The only information they learn is the result of the comparison.

\textbf{Security Parameters:}
We evaluate our approach with different public key security parameters $\varphi\in\{1024, 2048\}$ for bas-OTs. The symmetric security parameter $\kappa$ which determine the number of base-OTs is set to 80 or 128.

\textbf{Dataset:}
We evaluate our proposed approach using a genome database released by ``iDASH Security and Privacy Workshop 2016" \cite{IDASH}. Briefly, the server holds a database of 50 different sequences  and client holds one sequence to be evaluated against all the sequences in the database. The length of the sequences in average is 3500 characters from $\{A,C,G,T\}$ alphabet set. Therefore, the value of $n$ in 1-out-of-n OT would be 4.

\textbf{Setup:}
We evaluate our approach with respect to execution time and communication bandwidth. The goal of performance evaluation is to show the feasibility of \textbf{\textit{ESCOT}} protocol in real-world genome matching scenarios. We implement the framework in Java while client and server communicate through sockets. We run the framework over both LAN and WAN networks. For LAN setting, we use the VM machines provided by ``iDASH Security and Privacy Workshop 2016" so, we can provide a fair comparison with state of the art work \cite{al2017secure} which run the experiments on the same VM machines.  For WAN, we use an intercontinental cloud setting and perform the experiments on two free-tier Amazon instances with a 64-bit Intel Xeon dualcore CPU with 2.8 GHz and 3.75 GB RAM. The client and server are located in Oregon and Tokyo respectively. Evaluation over WAN gives us a better approximation of real-world scenarios. All the experiments are the average of 10 execution rounds.

\begin{figure*}
  
  \centering
    \includegraphics[width=0.8\textwidth]{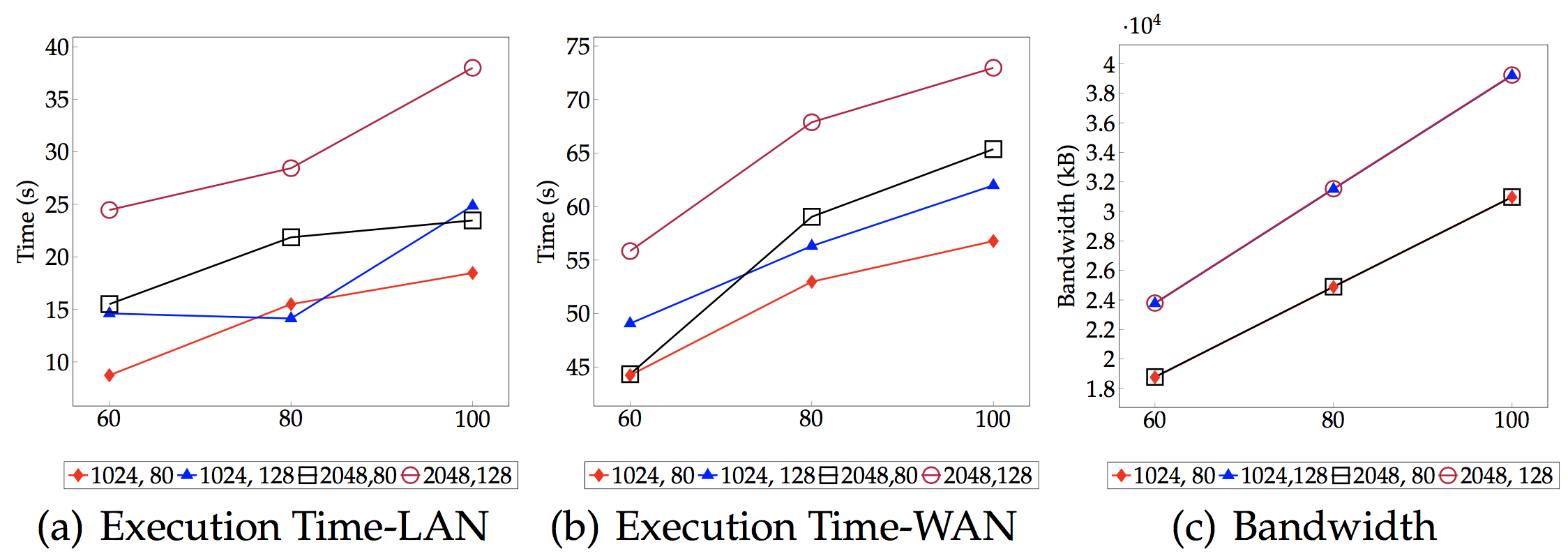}
    \caption{Privacy Preserving Edit Distance Performance Evaluation}\label{performance}
\end{figure*}

\textbf{Results:}
We set Edit distance threshold $k$ to 60, 80 and 100. The goal is to return the sequences with equal or less than the threshold $k$ distance to the client sequence. Obviously, by increasing the threshold the complexity $O(k \times m)$ increases. Experimental results are shown in Figure \ref{performance} with different security parameters ($\varphi\in\{1024, 2048\}$, $\kappa \in {80, 128}$). Figures (a) and (b) measures the running time in second on LAN and WAN network respectively while Figure (c)  shows  the bandwidth consumption in KB. Execution time varies from 8 to 38 seconds on LAN and 45 to 75 seconds on WAN. The execution time on WAN is higher due to network latency. Bandwidth directly depends on symmetric security parameter $\kappa$ or number of base-OTs, as it is shown in Figure (c) the public-key security parameter $\varphi$ does not affect the communication. The bandwidth varies from 18 to 40 MB.

\textbf{Analysis:} The private Edit distance protocols proposed in \cite{wang2015efficient} and \cite{al2017secure} are executed in 516 and 23 seconds respectively on the same dataset and over the same VM machines with baseline security parameters ($\varphi=1024$ and $\kappa=80$). While, our proposed protocol runs only in 8 seconds with same configuration. The other advantage of our approach over \cite{al2017secure} is that \textbf{\textit{ESCOT}} protocol calculates accurate Edit distance while the other work approximates the Edit distance. 

\subsection{Related Work}
Shantanu and Boufounos proposed an approach to calculate Edit distance using HE \cite{aguilar2013recent}. They reduced the problem to privacy-preserving minimum finding protocol that should be executed $m\times m'$ times ($m$ and $m'$ are the length of the input sequences). Huang et al. proposed a protocol to calculate Edit distance based on Garbled circuits \cite{huang2011faster}.

\section{Conclusion}
In this paper, we proposed an efficient solution for privacy preserving Edit distance using OT extension for the first time without losing accuracy. To do this, we proposed \textbf{\textit{ESCOT}} protocol for boolean comparison based on 1-out-of-n OT inspired by recent advances in secure two-party computation and Oblivious Transfer. We evaluate our approach on a genome dataset released by iDASH 2016 \cite{IDASH}. The experimental results confirm the efficiency of our approach over state of the art efforts for privacy preserving Edit distance.

\label{sect:bib}
\bibliographystyle{abbrv}
\bibliography{template}

\begin{thebibliography}{10}

\bibitem{IDASH}
In {\em GENOPRI WORKSHOP}.
  http://www.humangenomeprivacy.org/2016/competition-tasks.html, 2016.

\bibitem{aguilar2013recent}
C.~Aguilar-Melchor and et~al.
\newblock Recent advances in homomorphic encryption.
\newblock {\em Signal Processing Magazine}, 2013.

\bibitem{al2017secure}
M.~M. Al~Aziz and et~al.
\newblock Secure approximation of edit distance on genomic data.
\newblock {\em BMC Medical Genomics}, 2017.

\bibitem{asharov2013more}
G.~Asharov and et~al.
\newblock More efficient oblivious transfer and extensions for faster secure
  computation.
\newblock In {\em CCS}. ACM, 2013.

\bibitem{barni2010privacy}
M.~Barni and et~al.
\newblock Privacy-preserving fingercode authentication.
\newblock In {\em Multimedia and security}. ACM, 2010.

\bibitem{blanton2011secure}
M.~Blanton and P.~Gasti.
\newblock Secure and efficient protocols for iris and fingerprint
  identification.
\newblock In {\em ESORICS}. Springer, 2011.

\bibitem{bringer2012faster}
J.~Bringer and et~al.
\newblock Faster secure computation for biometric identification using
  filtering.
\newblock In {\em ICB}. IEEE, 2012.

\bibitem{cetin2017private}
G.~S. Cetin and et~al.
\newblock Private queries on encrypted genomic data.
\newblock {\em BMC Medical Genomics}, 2017.

\bibitem{chou2015simplest}
T.~Chou and C.~Orlandi.
\newblock The simplest protocol for oblivious transfer.
\newblock In {\em Cryptology and Information Security in Latin America}.
  Springer, 2015.

\bibitem{chun2014outsourceable}
H.~Chun and et~al.
\newblock Outsourceable two-party privacy-preserving biometric authentication.
\newblock In {\em ICCS}. ACM, 2014.

\bibitem{demmler2015aby}
D.~Demmler and et~al.
\newblock Aby-a framework for efficient mixed-protocol secure two-party
  computation.
\newblock In {\em NDSS}, 2015.

\bibitem{erkin2009privacy}
Z.~Erkin and et~al.
\newblock Privacy-preserving face recognition.
\newblock In {\em PETS}. Springer, 2009.

\bibitem{evans2011efficient}
D.~Evans and et~al.
\newblock Efficient privacy-preserving biometric identification.
\newblock In {\em NDSS}, 2011.

\bibitem{huang2011faster}
Y.~Huang and et~al.
\newblock Faster secure two-party computation using garbled circuits.
\newblock In {\em USENIX}, 2011.

\bibitem{kim2017secure}
M.~Kim and et~al.
\newblock Secure searching of biomarkers through hybrid homomorphic encryption
  scheme.
\newblock {\em BMC Medical Genomics}, 2017.

\bibitem{kolesnikov2013improved}
V.~Kolesnikov and R.~Kumaresan.
\newblock Improved ot extension for transferring short secrets.
\newblock In {\em CRYPTO}. Springer, 2013.

\bibitem{luo2012efficient}
Y.~Luo and et~al.
\newblock An efficient protocol for private iris-code matching by means of
  garbled circuits.
\newblock In {\em Image Processing}. IEEE, 2012.

\bibitem{BAYSlides}
T.~Schneider.
\newblock Aby - a framework for efficient mixed-protocol secure two-party
  computation.
\newblock In {\em Securing Computation Workshop}. EC SPRIDE, 2015.

\bibitem{ukkonen1985algorithms}
E.~Ukkonen.
\newblock Algorithms for approximate string matching.
\newblock {\em Information and control}, 1985.

\bibitem{wagner1975complexity}
R.~A. Wagner.
\newblock On the complexity of the extended string-to-string correction
  problem.
\newblock In {\em Theory of computing}. ACM, 1975.

\bibitem{wang2015efficient}
X.~S. Wang and et~al.
\newblock Efficient genome-wide, privacy-preserving similar patient query based
  on private edit distance.
\newblock In {\em CCS}. ACM, 2015.

\end{thebibliography}

\end{document}